\documentclass[lineno]{jfm}

\usepackage{graphicx}
\usepackage[justification=justified,width=\linewidth]{caption}
\usepackage{newtxtext}
\usepackage{newtxmath}
\usepackage{natbib}
\usepackage{xcolor}

\definecolor{color_symbol_I}{rgb}{0.8500, 0.3250, 0.0980}
\definecolor{color_symbol_II}{rgb}{0.6350, 0.0780, 0.1840}
\definecolor{color_symbol_III}{rgb}{0, 0.4470, 0.7410}
\definecolor{color_symbol_IV}{rgb}{0.4660, 0.6740, 0.1880}
\definecolor{color_symbol_V}{rgb}{0, 0.5, 0}
\definecolor{color_symbol_VI}{rgb}{0.5 0.5 0.5}
\definecolor{color_symbol_VII}{rgb}{0.6, 0.2980, 0}

\newcommand{\RomanNumeralCaps}[1]
\linenumbers

\title{Non-linear regimes of tsunami waves generated by a granular collapse}

\author{Wladimir Sarlin\aff{1},
  Cyprien Morize\aff{1}
    \corresp{\email{cyprien.morize@universite-paris-saclay.fr}},
  Alban Sauret\aff{2}
 \and Philippe Gondret\aff{1}}

\affiliation{\aff{1}Universit\'e Paris-Saclay, CNRS, Laboratoire FAST, F-91405 Orsay, France
\aff{2} University of California, Santa Barbara, Department of Mechanical Engineering, CA 93106, USA}

\begin{document}
\maketitle

\begin{abstract}
Tsunami waves induced by landslides are a threat to human activities and safety along coastal areas. In this paper, we characterize experimentally the waves generated by the gravity-driven collapse of a dry granular column into water. Three nonlinear wave regimes are identified depending on the Froude number $\mathrm{Fr}_f$ based on the ratio of the velocity of the advancing granular front and the velocity of linear gravity waves in shallow water: transient bores for large $\mathrm{Fr}_f$, solitary waves for intermediate values of $\mathrm{Fr}_f$, and non-linear transition waves at small $\mathrm{Fr}_f$. The wave amplitude relative to the water depth increases with $\mathrm{Fr}_f$ in the three regimes but with different non-linear scalings, and the relative wavelength is an increasing or decreasing function of $\mathrm{Fr}_f$. Two of these wave regimes are rationalized by considering that the advancing granular front acts as a vertical piston pushing the water, while the last one is found to be a transition from shallow to deep water conditions. The present modeling contributes to a better understanding of the rich hydrodynamics of the generated waves, with coastal risk assessment as practical applications.
\end{abstract}

\begin{keywords}
solitary waves, surface gravity waves, avalanches
\end{keywords}

\section{Introduction}
\label{sec:1}

In 2018, the partial flank  collapse of Anak Krakatau led to a tsunami that caused major human casualties and material damage along the neighbouring coast \citep{Grilli2019, Paris2020}. Many other volcanic islands are susceptible to a similar collapse with an associated risk of tsunamis, such as La R\'eunion in the Indian Ocean \citep{Kelfoun2010} or La Palma in the Atlantic Ocean \citep{Abadie2012}. The generation of tsunami waves by landslides may be triggered by volcanic or seismic events not only in the ocean, but also in lakes or rivers \citep{Kremer2012, Couston2015}, due to the collapse or avalanche of either soil, rocks, or even ice and snow \citep{Zitti2016}. The experimental and numerical studies of \citet{Clous2019} and \citet{Cabrera2020} have recently shown that subaerial landslides trigger much larger waves than submarine landslides for a given amount of destabilized materials.

For subaerial events, the Froude number, corresponding to the ratio of the velocity of the landslide entering into the water to the wave velocity, is expected to play a crucial role. The simplest approach to model the generation of a tsunami wave by a landslide is to consider the impact of a sliding wedge on an inclined plane \citep{2003_walder}. However, granular materials must be considered to account for the complex landslide motion and interplay with the water. By studying the entry of grains into water at high velocity from a pneumatically launched box along a smooth inclined plane, \citet{2004_fritz} observed different wave regimes depending on the Froude number Fr and slide thickness $S$ : (i) transient (breaking) bores at high Fr and $S$; (ii) solitary-like waves at moderate Fr and $S$; (iii) non-linear transition waves at low Fr and $S$; and (iv) weakly non-linear oscillatory waves at very low Fr and $S$. \citet{2017_robbe-saule} have considered the experimental gravity-driven collapse of a subaerial granular column into water. They have shown that the aspect ratio and the volume of the granular column both play an important role on the amplitude of the wave generated. With a similar but larger set-up, \citet{Huang2020} observed the first three wave regimes reported by \citet{2004_fritz}, again depending on the aspect ratio and volume of the column. Recently, \citet{2021_robbe-saule} have shown that the local Froude number $\mathrm{Fr}_f$ based on the horizontal velocity $v_f$ of the granular front at the water surface is the relevant parameter that governs the generation of the wave. However, a theoretical framework that accounts for the different wave regimes observed experimentally remains elusive. This lack of knowledge makes difficult the development of accurate predictive models and coastal risk assessment in the context of tsunamis generated by landslides, which is one of the grand challenges in environmental fluid mechanics \citep{Dauxois2021}.

In this paper, we report experimental results of the wave generated by a gravity-driven granular collapse into water for a large range of local Froude number $\mathrm{Fr}_f$, and characterize the three wave regimes observed : (i) transient bore waves at high $\mathrm{Fr}_f$; (ii) solitary waves in intermediate $\mathrm{Fr}_f$; and (iii) non-linear transition waves at small $\mathrm{Fr}_f$. For the first two regimes, theoretical models from the shallow-water wave equations are then developed, which compare well to the experimental results.

\section{Experimental setup and results}
\label{sec:2}

\begin{figure}
  {\includegraphics[width=\textwidth]{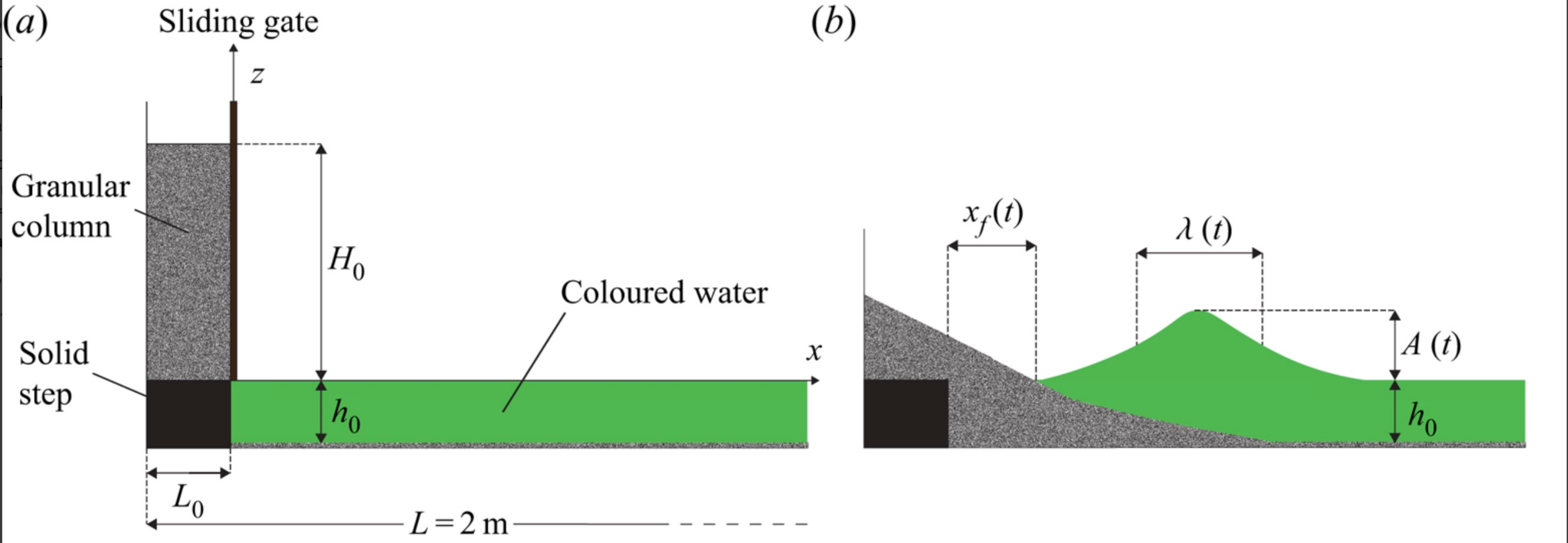}}  
  \caption{Sketch of the experimental set-up (a) in the initial configuration, with a dry granular column of height $H_0$ and length $L_0$ above a water depth $h_0$, and (b) during the granular collapse, with an advancing front $x_f(t)$ at the water surface and a generated wave of amplitude $A(t)$ and mid-height width $\lambda(t)$.}
\label{fig:fig_I}
\end{figure}

\begin{table}
  \begin{center}
\def~{\hphantom{0}}
  \begin{tabular}{ccccccc}
 $H_{0}$ (cm) & $L_{0}$ (cm) & $V_{0}$ (dm$^3$) & $a$ & $h_{0}$ (cm) & $\mathrm{Fr}_0$ & Symbols \\
 9 & 10 & 1.35 & 0.9 & 2 -- 25 & 0.6 -- 2.1 & {\footnotesize $ \textcolor{color_symbol_I}{\blacksquare} $}\\
 19 & 10 & 2.85 & 1.9 & 2 -- 25 & 0.9 -- 3.1 & {\Large $ \textcolor{color_symbol_II}{\bullet} $}\\
 29 & 10 & 4.35 & 2.9 & 2 -- 20 & 1.2 -- 3.8 & $ \textcolor{color_symbol_III}{\blacktriangle} $\\
 39 & 5 & 2.93 & 7. 8 & 2 -- 25 & 1.2 -- 4.4 & $ \textcolor{color_symbol_IV}{\blacktriangledown}$\\
 39 & 10 & 5.85 & 3.9 & 2 -- 20 & 1.4 -- 4.4 & $ \textcolor{color_symbol_V}{\bigstar} $\\
 39 & 14.5 & 8.48 & 2.7 & 2 -- 20 & 1.4 -- 4.4 & $ \textcolor{color_symbol_VI}{\blacklozenge} $\\
 39 & 20 & 11.7 & 2.0 & 2 -- 20 & 1.2 -- 4.4 & $ \textcolor{color_symbol_VII}{\blacktriangleright} $\\
  \end{tabular}
  \caption{Sets of experimental parameters  with corresponding data symbols.}
  \label{tab:table_I}
  \end{center}
\end{table}

\subsection{Set-up}
\label{subsec:2.1}

We perform new experiments using the two-dimensional set-up illustrated in figure \ref{fig:fig_I}(a) and described in detail in \citet{2021_robbe-saule}.
A rectangular tank of length $L=2$~m and transverse width $W=0.15$~m is filled up to a height $h_0$ with initially still water. On the left side of the tank, a rectangular granular column of height $H_0$ and length $L_0$ stands on a solid step of height $h_0$ so that the grains, initially retained by a vertical gate, are just above the water interface. The granular material consists of monodisperse glass beads of diameter $5$~mm and density $\rho =$ 2.5 g/cm$^3$. Images are taken by a camera from the sidewall of the tank. The water is dyed with fluorescein to enhance the contrast and facilitate the processing of the time evolution of the free surface of the water and of the grains. \\
At time $t=0$, the gate is quickly lifted by a linear motor at 1 m.s$^{-1}$. The granular column then collapses into the water leading to an advancing granular front $x_f(t)$, which generates an impulse wave of amplitude $A(t)$ and mid-height width $\lambda(t)$, as sketched in figure \ref{fig:fig_I}(b). We perform systematic experiments, where both the size of the column and the water depth are varied, as detailed in table \ref{tab:table_I}. We explore a large range of aspect ratio $a=H_0/L_0$, initial volume of the column $V_0=H_0L_0W$, and global Froude number $\mathrm{Fr}_0=\sqrt{H_0/h_0}$, which compares the typical vertical free-fall velocity of the granular medium, $\sqrt{gH_0}$, to the velocity of linear gravity waves in shallow water, $c_0=\sqrt{gh_0}$.

\subsection{Description of the observed regimes of non-linear waves}
\label{subsec:2.2}

\begin{figure}
{\includegraphics[width=\linewidth]{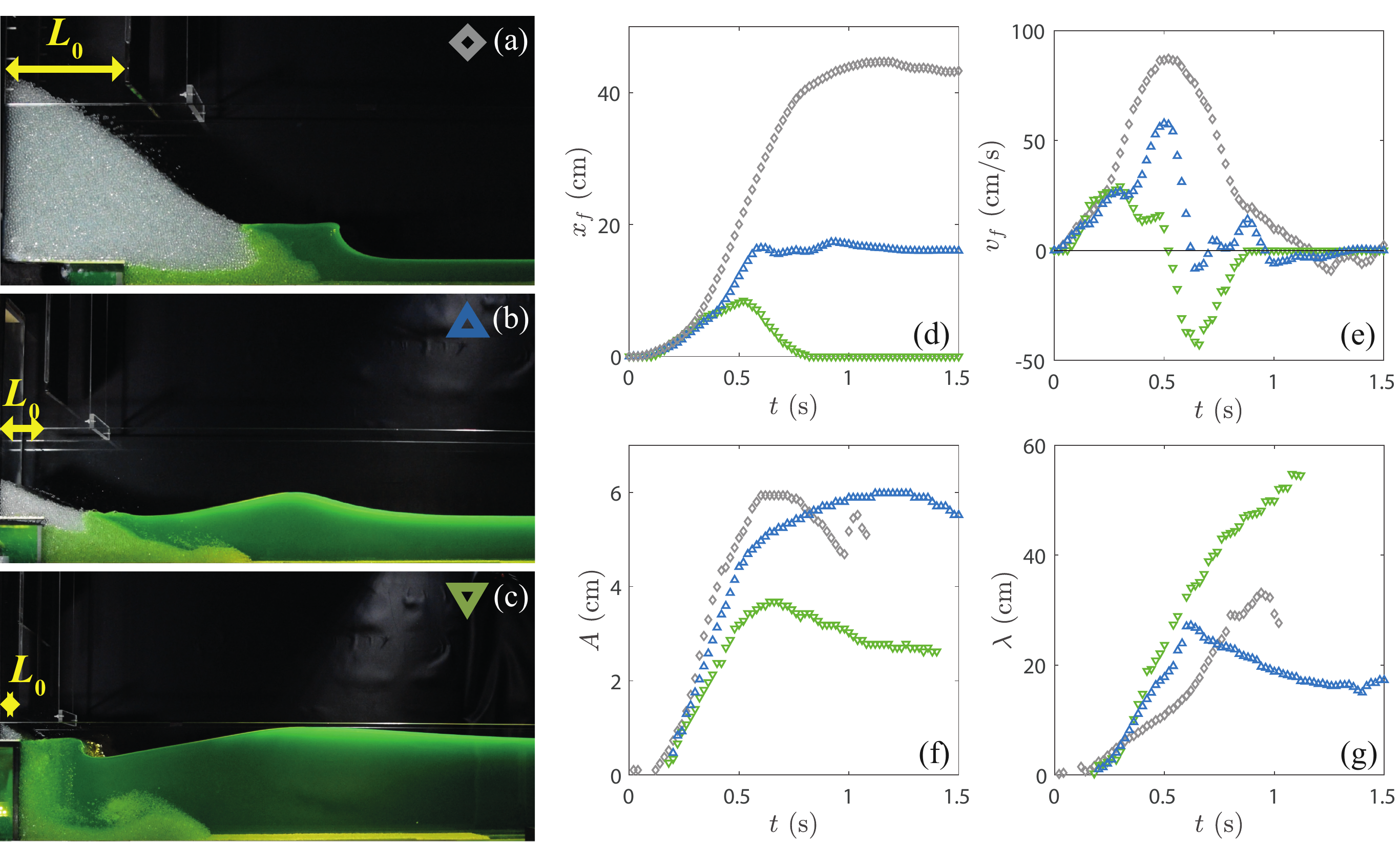}}
\caption{(a-c) Photographs of the different wave shapes observed : (a) bore wave at $H_0$ = 39 cm, $L_0$ = 14.5 cm, $h_0$ = 4 cm ($\mathrm{Fr}_0=3.1$, $\mathrm{Fr}_f=1.39$); (b) solitary wave at $H_0$ = 29 cm, $L_0$ = 10 cm, $h_0$ = 8 cm ($\mathrm{Fr}_0=1.9$, $\mathrm{Fr}_f=0.65$); and (c) nonlinear transition wave at $H_0$ = 39 cm, $L_0$ = 5 cm, $h_0$ = 25 cm ($\mathrm{Fr}_0=1.2$, $\mathrm{Fr}_f=0.19$). (d-g) Time series of (d) the position $x_f$ and (e) the velocity $v_f$ of the granular front, and of (f) the amplitude $A$ and (g) the mid-height width $\lambda$ of the wave for the experiments of panels (a) (\textcolor{color_symbol_VI}{\Large $\diamond$}), (b) (\textcolor{color_symbol_III}{\large $\triangle$}), and (c) (\textcolor{color_symbol_IV}{\large $\triangledown$}).}
\label{fig:fig_II}
\end{figure}

Depending on the geometry of the granular columns and the water depth, leading to different global Froude number $\mathrm{Fr}_0$, we observe three different regimes of non-linear waves [see figures \ref{fig:fig_II}(a)-(c), and videos in supplementary materials]. At large $\mathrm{Fr}_0$, strong asymmetric waves are generated with the shape of transient positive surges or bores, as shown in figure \ref{fig:fig_II}(a). The wave generated systematically breaks in the near-field region, and corresponds to the plunging breaker reported in \citet{2021_robbe-saule}. At moderate $\mathrm{Fr}_0$, quasi-symmetrical waves are generated, consisting of a unique main pulse of soliton-like shape, as reported in figure \ref{fig:fig_II}(b). This wave may break or not depending on its relative amplitude $A/h_0$. The breaking cases at high $A/h_0$ correspond to spilling breakers \citep{2021_robbe-saule}. At low $\mathrm{Fr}_0$, waves with slightly reversed asymmetry and strong unsteadiness are generated [see figure \ref{fig:fig_II}(c)]. This situation corresponds to the non-linear transition waves reported by \citet{2004_fritz} and \citet{2013_viroulet}.
\\Figures \ref{fig:fig_II}(d)-(f) report the time evolution of the granular collapse and of the generated wave for the three examples of figures \ref{fig:fig_II}(a)-(c).
At high $\mathrm{Fr}_0$ (\textcolor{color_symbol_VI}{\LARGE $\diamond$}), the position of the granular front, $x_{f}(t)$, continually increases from zero to a final maximum value $x_{f_\infty} \gg h_0$ ($x_{f_\infty} \simeq 11 h_0$ in figure \ref{fig:fig_II}(d)). The corresponding velocity of the front $v_f=\mathrm{d} x_{f}/\mathrm{d} t$ exhibits a bell shape from zero up to a maximum value $v_{f_m} \gtrsim \sqrt{gh_0}$ ($v_{f_m} \simeq 1.4 \sqrt{gh_0}$ in figure \ref{fig:fig_II}(e)) and then decreases to zero. Both the amplitude $A$ and the mid-height width $\lambda$ increase during the generation process, until reaching a maximum value $A_m \gtrsim h_0$ with the corresponding $\lambda_m \gg h_0$ ($A_m \simeq 1.5 h_0$ and $\lambda_m \simeq 5 h_0$ in figures \ref{fig:fig_II}(f) and \ref{fig:fig_II}(g), respectively), at the moment where the plunging breakup occurs ($t \simeq 0.7$s in figure \ref{fig:fig_II}(e)), which leads to a sudden decrease~of~$A$.\\
At moderate $\mathrm{Fr}_0$ (\textcolor{color_symbol_III}{\Large $\triangle$}), the time evolution of the granular collapse is similar to that described above but with a smaller maximal extension $x_{f_\infty} \gtrsim h_0$ and a smaller maximum velocity $v_{f_m} \simeq \sqrt{gh_0}$ ($x_{f_\infty} \simeq  2h_0 $ and $v_{f_m} \simeq 0.7 \sqrt{gh_0}$ in figures \ref{fig:fig_II}(d) and \ref{fig:fig_II}(e), respectively). The wave amplitude first increases abruptly and then more slowly to a maximum value $A_m  \simeq h_0$ ($A_m  \simeq 0.75 h_0$ in figure \ref{fig:fig_II}(f)) before decreasing due to the breaking of the wave. In the meantime, the width $\lambda$ first grows up to a maximum value, and then decreases down to $\lambda_m > h_0$ ($\lambda_m  \simeq 3 h_0$ in figure \ref{fig:fig_II}(f)) when the wave propagates away from the collapse.\\
At low $\mathrm{Fr}_0$ (\textcolor{color_symbol_IV}{\Large $\triangledown$}), the time evolution of the granular collapse is very different from the two previous situations : $x_{f}$ increases from zero up to a maximum value $x_{f_m} \lesssim h_0$ ($x_{f_m} \simeq 0.3 h_0$ in figure \ref{fig:fig_II}(d)) before decreasing down to zero when the flow rate of the granular medium starts to vanish. In this configuration, the bottom wall does not play a significant role in the granular dynamics. The corresponding velocity $v_{f}(t)$ exhibits first a positive maximum $v_{f_m} < \sqrt{gh_0} $ ($v_{f_m} \simeq 0.2 \sqrt{gh_0}$ in figure \ref{fig:fig_II}(e)), but also a negative minimum at the end of the collapse, corresponding to the receding phase of the granular flow. In the meantime, while the wave amplitude first increases to a maximum value $A_m  < h_0$ before slightly decreasing, the mid-height width displays a monotonic growth beyond $\lambda_m \simeq  h_0$ ($A_m \simeq 0.3 h_0$ and $\lambda_m \simeq 1.2 h_0$ in figures \ref{fig:fig_II}(f) and \ref{fig:fig_II}(g), respectively), revealing that the wave flattens after its generation.
These waves all exhibit a wavelength much larger than the capillary length $\lambda_c \simeq 1.6$ cm and larger than the water depth (except for a few experiments that are discussed in subsection \ref{subsec:3.1}) so that they correspond to gravity waves in shallow-water conditions.

\begin{figure}
  {\includegraphics[width=0.5\linewidth]{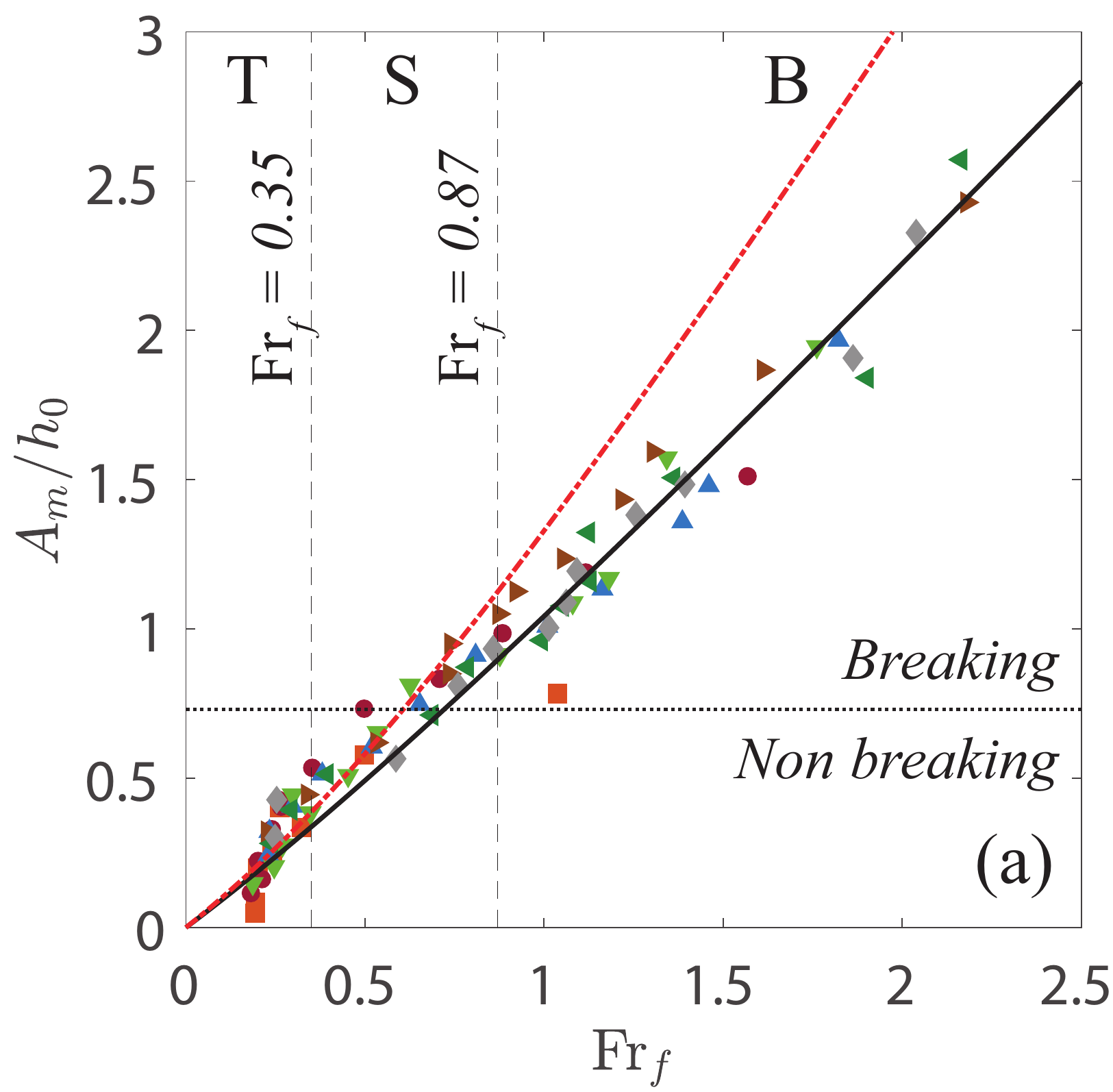}}
  \hfill
  {\includegraphics[width=0.5\linewidth]{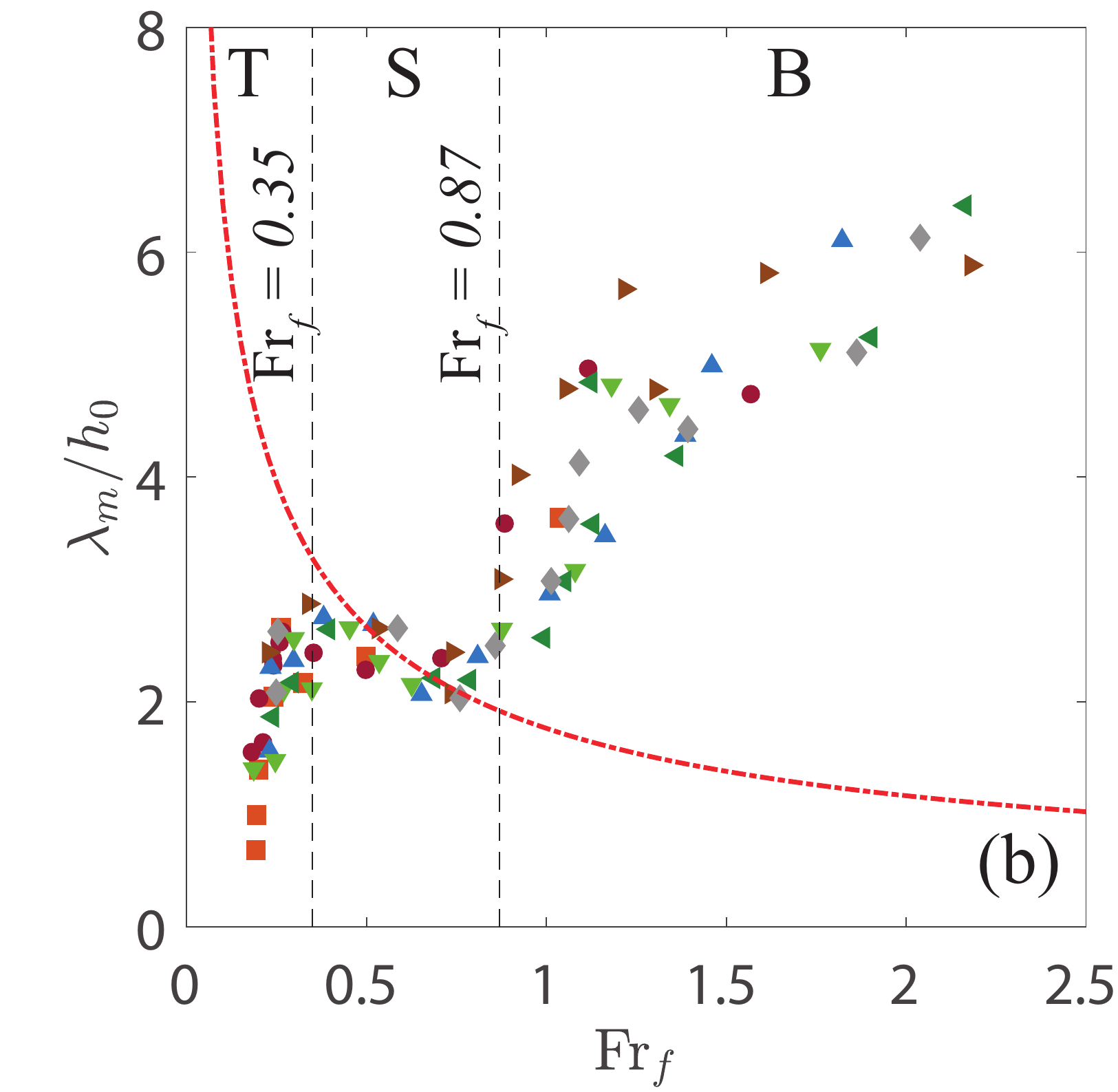}}
  \caption{(a) Relative amplitude $A_m/h_0$ and (b) mid-height width $\lambda_m/h_0$ of the generated waves as functions of the local Froude number $\mathrm{Fr}_f$ for all the experiments of table \ref{tab:table_I}, with predictions from equations (\ref{eqArelHJ}) (---) and (\ref{eqArelSW}) ({\color{red} --.--}). The two vertical thin dashed lines at $\mathrm{Fr}_f = 0.35$ and $\mathrm{Fr}_f = 0.87$ separate the three regimes corresponding to nonlinear transition waves (T), solitary waves (S) and bore waves (B), respectively. The horizontal dotted line at $A_m/h_0 \simeq 0.73$ separates breaking and non-breaking waves.}
\label{fig:fig_III}
\end{figure}

The local Froude number, $\mathrm{Fr}_f = v_{f_m}/\sqrt{gh_0}$, based on the velocity of the granular front, was found to be the relevant dimensionless parameter to describe the wave generation \citep{2021_robbe-saule}. All the experimental results for the maximum wave amplitude $A_m$ and the associated mid-height width $\lambda_m$, non-dimensionalized by the water depth $h_0$, are plotted as a function of  $\mathrm{Fr}_f$ in figure \ref{fig:fig_III}(a)-(b). A monotonic increase of the relative wave amplitude $A_m/h_0$ with $\mathrm{Fr}_f$ is observed, whereas the relative mid-height width $\lambda_m/h_0$ first increases but then slightly decreases before increasing again. These different behaviours lead to a clear separation of the three wave regimes described above : non-linear transition waves (T) for $\mathrm{Fr}_f \lesssim 0.35$, solitary waves (S) for $0.35 \lesssim \mathrm{Fr}_f \lesssim 0.87$, and transient bore waves (B) for $\mathrm{Fr}_f \gtrsim 0.87$. In the next section, we characterize the wave regimes using theoretical models adapted to the different regimes and obtain scaling laws for the maximal wave amplitude and the corresponding wavelength with the local Froude number $\mathrm{Fr}_f$, for regimes S and B.

\section{Modelling the generation of the different non-linear waves}
\label{sec:3}

\subsection{Bore waves}
\label{subsec:3.3}

For large values of $\mathrm{Fr}_f$ (regime B, $\mathrm{Fr}_f \gtrsim 0.87$), the deformation of the free surface of the water is very similar to a bore or a positive surge during the generation process : the shape is an elongated water bump of height $h_0 + A_B$ with an abrupt step front propagating over the downstream region of height $h_0$. These waves are observed during the collapse of granular columns with a large initial height $H_0$ relative to the water depth $h_0$, and therefore for large $\mathrm{Fr}_0$. In such conditions, the advancing granular front is almost vertical throughout the entire water depth and acts as a vertical translating piston over a distance $x \gg h_0$ at the velocity $v_p$ [see figure \ref{fig:fig_II}(a)]. We consider the mass and momentum conservation equations in the frame of reference of the bore propagating at the velocity $c$,

\begin{figure}
  \centering
  {\includegraphics[width=0.48\linewidth]{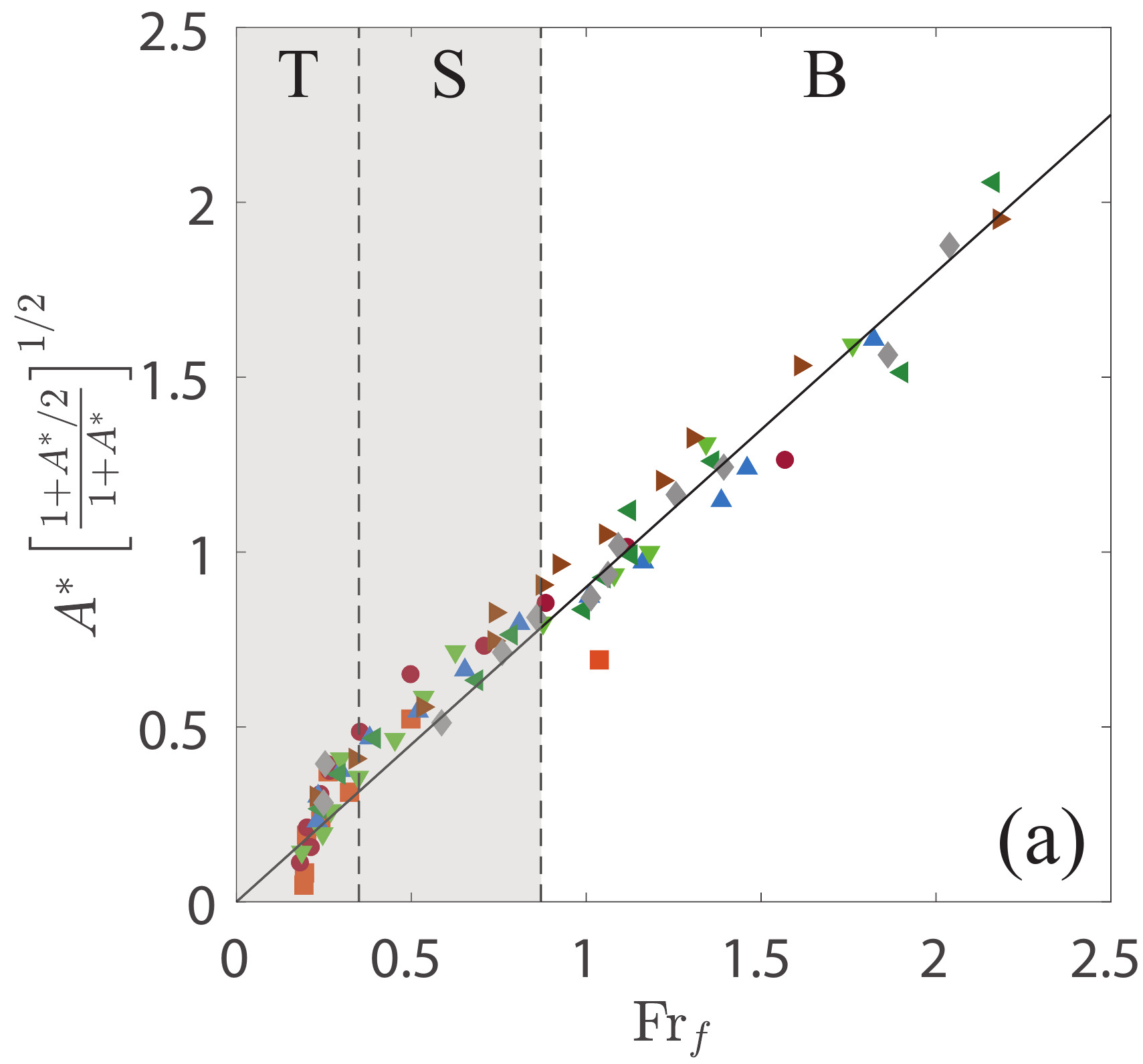}}
  {\includegraphics[width=0.48\linewidth]{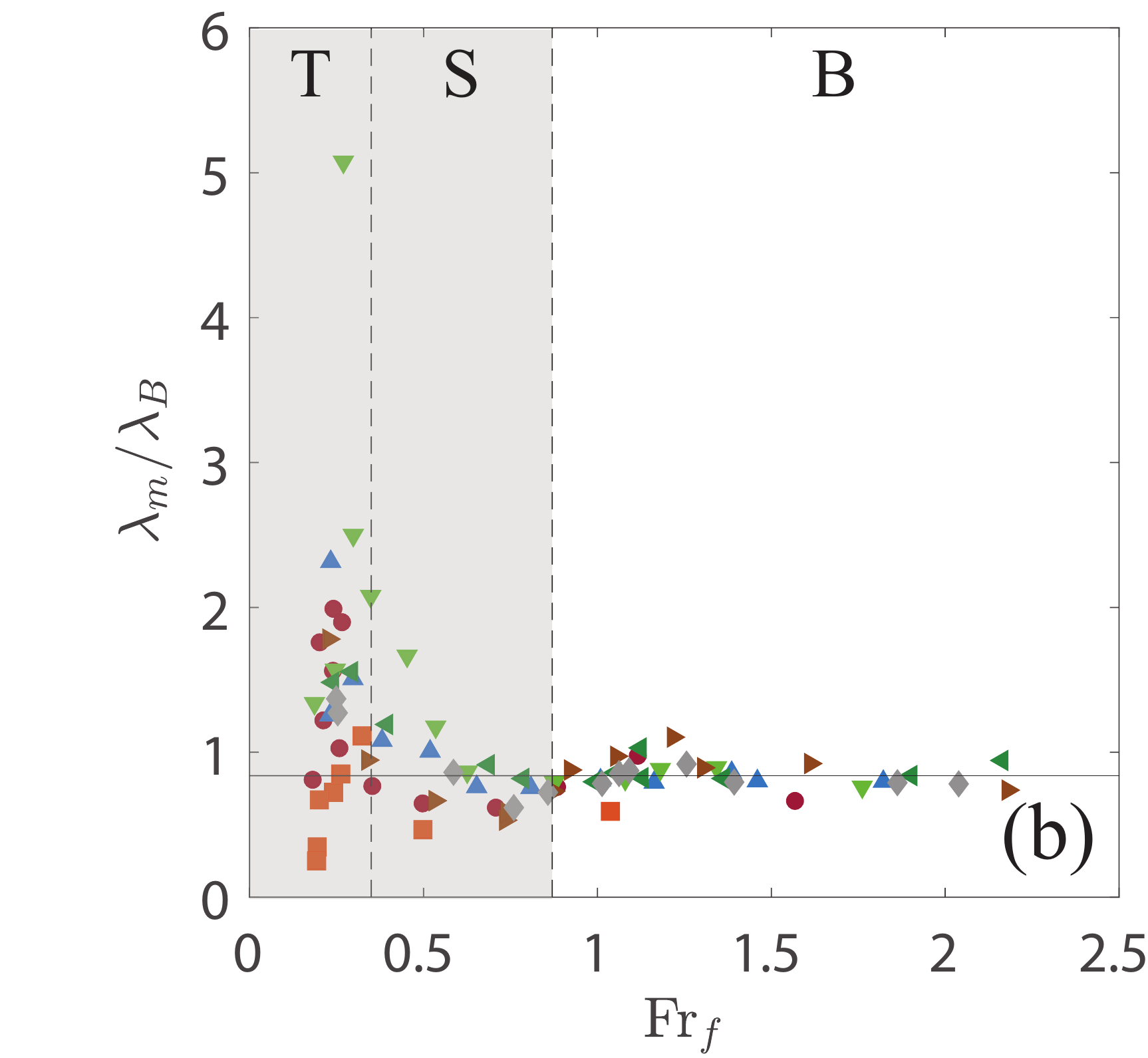}}
  \caption{(a) Plot of $A^*[(1+A^*/2)/(1+A^*)]^{1/2}$ as a function of $\mathrm{Fr}_{f}$ for all experimental data of table \ref{tab:table_I}, with $A^*=A_m/h_0$, and the best linear fit (full line) with a slope 0.9 for the data corresponding to bore waves ($\mathrm{Fr}_f \gtrsim 0.87$). (b) Ratio of the experimental mid-height width $\lambda_m$ of the wave to the calculated one, $\lambda_B$, from equation (\ref{eqlambdamHJ}), as a function of $\mathrm{Fr}_f$. The horizontal solid line corresponds to $\lambda_m/\lambda_B=0.84$.}
\label{fig:fig_IV}
\end{figure}

\begin{equation}
	\left\{ \begin{array}{ll}
	\displaystyle (c-v_p) (h_0+A_B)-c h_0 = 0,\\[8pt]
	\displaystyle  (c-v_p)^2(h_0 + A_B)-c^2h_0+\frac{g}{2} \left[ (h_0 + A_B)^2-h_0^2 \right] =0.
	\end{array}\right.
  \label{eqconseqs}
\end{equation}

\noindent These equations describe a stationary hydraulic jump of amplitude $A_B$, separating a thin supercritical region of depth $h_0$ and constant velocity $-c$ from a thick subcritical region of depth $h_0 + A_B$ and of constant velocity $v_p-c$, when assuming no dissipation at the bottom wall and a hydrostatic vertical pressure gradient far enough from the jump \citep{Guyon2015}. Combining these two equations leads to a non-linear relation between the relative amplitude $A_B/h_0$ and the Froude number $\mathrm{Fr}_p = v_p/\sqrt{gh_0}$ based on the piston velocity~$v_p$ :

\begin{equation}
\frac{A_B}{h_0} \left( \frac{1+A_B/(2h_0)}{1+A_B/h_0} \right)^{1/2}= \mathrm{Fr}_p.
  \label{eqArelvsFrHJ}
\end{equation}

\noindent Figure \ref{fig:fig_IV}(a) shows the rescaled experimental data using the left side of equation (\ref{eqArelvsFrHJ}), with $A_B=A_m$, as a function of $\mathrm{Fr}_f$. The data corresponding to $\mathrm{Fr}_f \gtrsim 0.87$ collapse onto a straight line of slope 0.9. Therefore, these nonlinear waves can be seen as bores generated by a solid wall pushing the water at an effective constant velocity $v_p$, a little smaller than the maximal velocity of the granular front $v_{f_m}$. The explicit expression for $A_B/h_0$ as a function of $\mathrm{Fr}_p$ can be obtained as the only positive solution of the third-order equation  (\ref{eqArelvsFrHJ}):

\begin{equation}
  \frac{A_B}{h_{0}}=\frac{2}{3} \left[ 2 \sqrt{ 1 + 3{\mathrm{Fr}_p}^2 /2 } \cos \left( \frac{1}{3} \cos^{-1} \left[ \frac{3}{4}\frac{(\mathrm{Fr}_p-2\sqrt{2}/3)(\mathrm{Fr}_p+2\sqrt{2}/3)}{({\mathrm{Fr}_p}^2+2/3) \sqrt{ 1 + 3{\mathrm{Fr}_p}^2 /2 }} \right] \right) - 1 \right].
  \label{eqArelHJ}
\end{equation}

\noindent The solid line in figure \ref{fig:fig_III}(a) corresponds to equation (\ref{eqArelHJ}) with $\mathrm{Fr}_p = 0.9 \, \mathrm{Fr}_f$. This prediction fits very well the data for $\mathrm{Fr}_f \gtrsim 0.87$, i.e. in regime B. Note that this non-linear variation would asymptotically tend to the linear expression $A_B/h_0 \sim  \sqrt{2} \, \mathrm{Fr}_p$ as $\mathrm{Fr}_p \rightarrow +\infty$.
The wavelength $\lambda_B$ can also be estimated from the mass conservation at the end of the generation of the bore, assuming that the hydraulic jump has the shape of a pure step. Indeed, when the piston has travelled over the total distance $x_p$, we should have

\begin{equation}
\lambda_B=x_p \frac{h_0}{A_B}.
  \label{eqlambdamHJ}
\end{equation}

\noindent Figure \ref{fig:fig_IV}(b) presents the ratio $\lambda_m/\lambda_B$ as a function of $\mathrm{Fr}_f$. Here, we consider that $x_p$ corresponds to the value of $x_f$ at the end of the wave generation, when $A=A_m$. A plateau is observed with a value $\lambda_m/\lambda_B \simeq 0.84$ for $\mathrm{Fr}_f \gtrsim 0.87$. Since the hydraulic jump is not a pure step and the granular front is not exactly a vertical advancing wall at a constant velocity, the observed wavelength $\lambda_m$ is indeed a little smaller than the predicted value $\lambda_B$. Note that equation (\ref{eqlambdamHJ}) implies that $\lambda_B/h_0 \sim (x_{p}/h_0){\mathrm{Fr}_f}^{-1}$ at large $\mathrm{Fr}_f$, where $\lambda_B/h_0$ depends not only on $\mathrm{Fr}_f$ but also on $x_{p}/h_0$. This explains why the experimental data $\lambda_m/h_0$ of regime B do not collapse onto a master curve in figure \ref{fig:fig_III}(b), but are quite dispersed, as they also depend on $x_{f_m}/h_0$.

\subsection{Solitary waves}
\label{subsec:3.2}

In an intermediate range of local Froude number (regime S, $ 0.35 \lesssim \mathrm{Fr}_f \lesssim 0.87$), the generated waves exhibit a solitary-like shape. Solitons are solutions of the Korteweg-de Vries equation valid for shallow-water waves without dissipation \citep{2006_dauxois}. In that theoretical framework, the free surface elevation $\eta(x,t)$ of a soliton is given by

\begin{equation}
	\eta(x,t)=A_S \operatorname{sech}^2 \left( \frac{c_{S}t-x}{ \lambda_{S} } \right),\ \mathrm{with}\ \lambda_{S} = 2h_0\sqrt{\frac{h_0}{3A_S}},\ \mathrm{and}\ c_{S}=\sqrt{gh_0} \left[ 1+\frac{A_S}{2h_0} \right],
  \label{eqKdVtheory}
\end{equation}

\noindent where $A_S$, $\lambda_{S}$ and $c_{S}$ are the amplitude, characteristic width and velocity of the wave, respectively, and $\operatorname{sech}$ is the hyperbolic secant function. These solitons can be generated experimentally by wavemakers with a vertical piston moving according to the following law \citep{1980_goring,1990_synolakis,2002_guizien}:

\begin{equation}
	x_p(t)=\frac{A_S\lambda_{S}}{h_0}\tanh \left( \frac{c_{S}t-x_p(t)}{\lambda_{S}} \right).
  \label{eqmotion}
\end{equation}

\noindent Note that, for the experiments in this regime, the time evolution of the granular front $x_f(t)$ is close to a hyperbolic tangent evolution, as shown by the curve in figure \ref{fig:fig_II}(d). The maximum value of the time derivative of equation (\ref{eqmotion}) leads to the following equation for the maximum velocity $v_p$ of the piston:

\begin{equation}
v_p=\frac{A_S}{h_0}(c_{S}-v_p).
\label{eqpistonmaxspeed}
\end{equation}

\noindent Considering the expression (\ref{eqKdVtheory}c) for $c_{S}$ together with equation (\ref{eqpistonmaxspeed}) leads to a relation between the relative wave amplitude of the solitary wave $A_S/h_0$ and the Froude number $\mathrm{Fr}_p$ based on the piston velocity $v_p$ :

\begin{equation}
\frac{A_S}{h_0} \frac{1+A_S/(2h_0)}{1+A_S/h_0}=\mathrm{Fr}_p.
\label{eqArelSWint}
\end{equation}

\begin{figure}
  \centering
  {\includegraphics[width=0.48\linewidth]{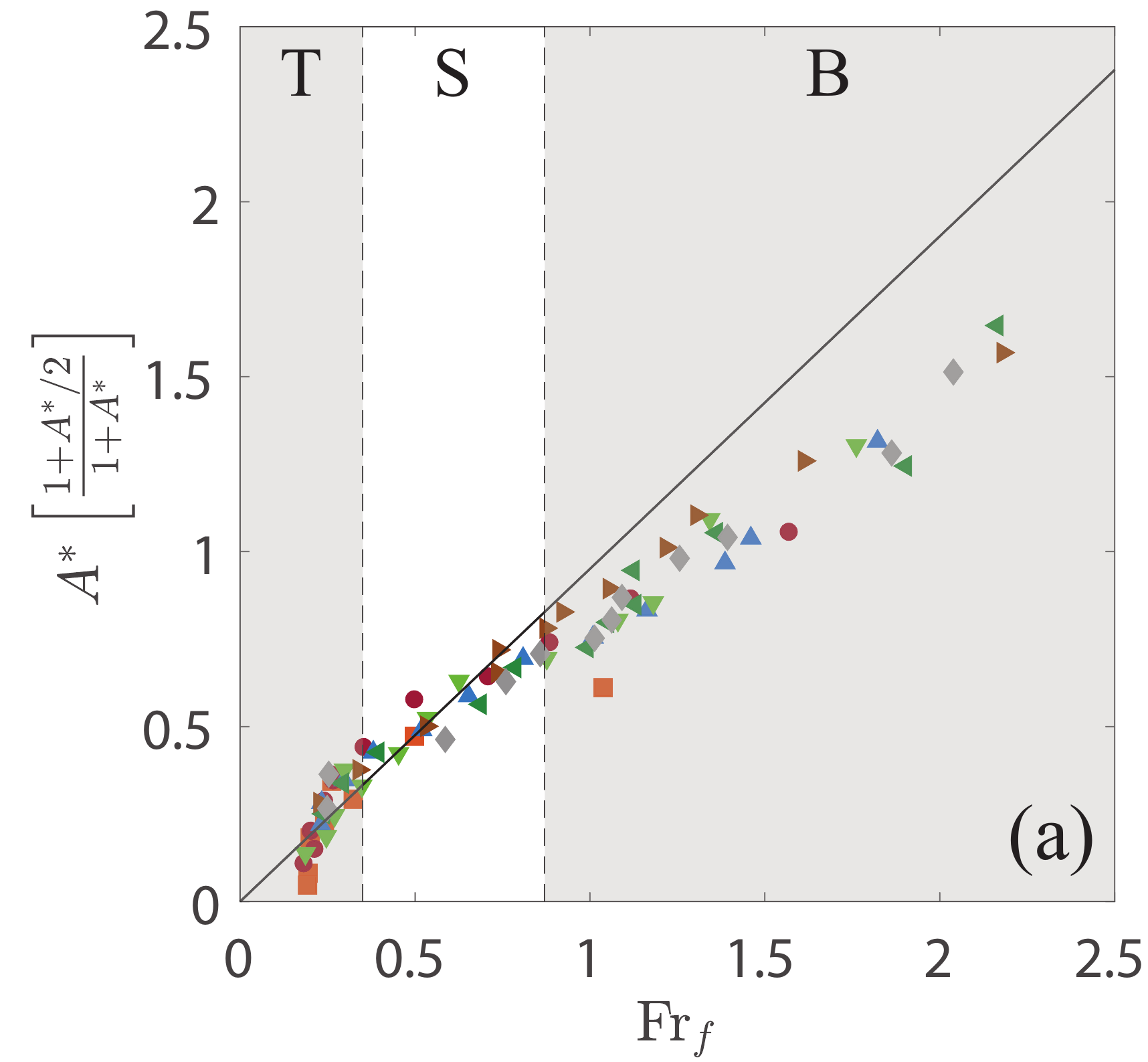}}
  {\includegraphics[width=0.48\linewidth]{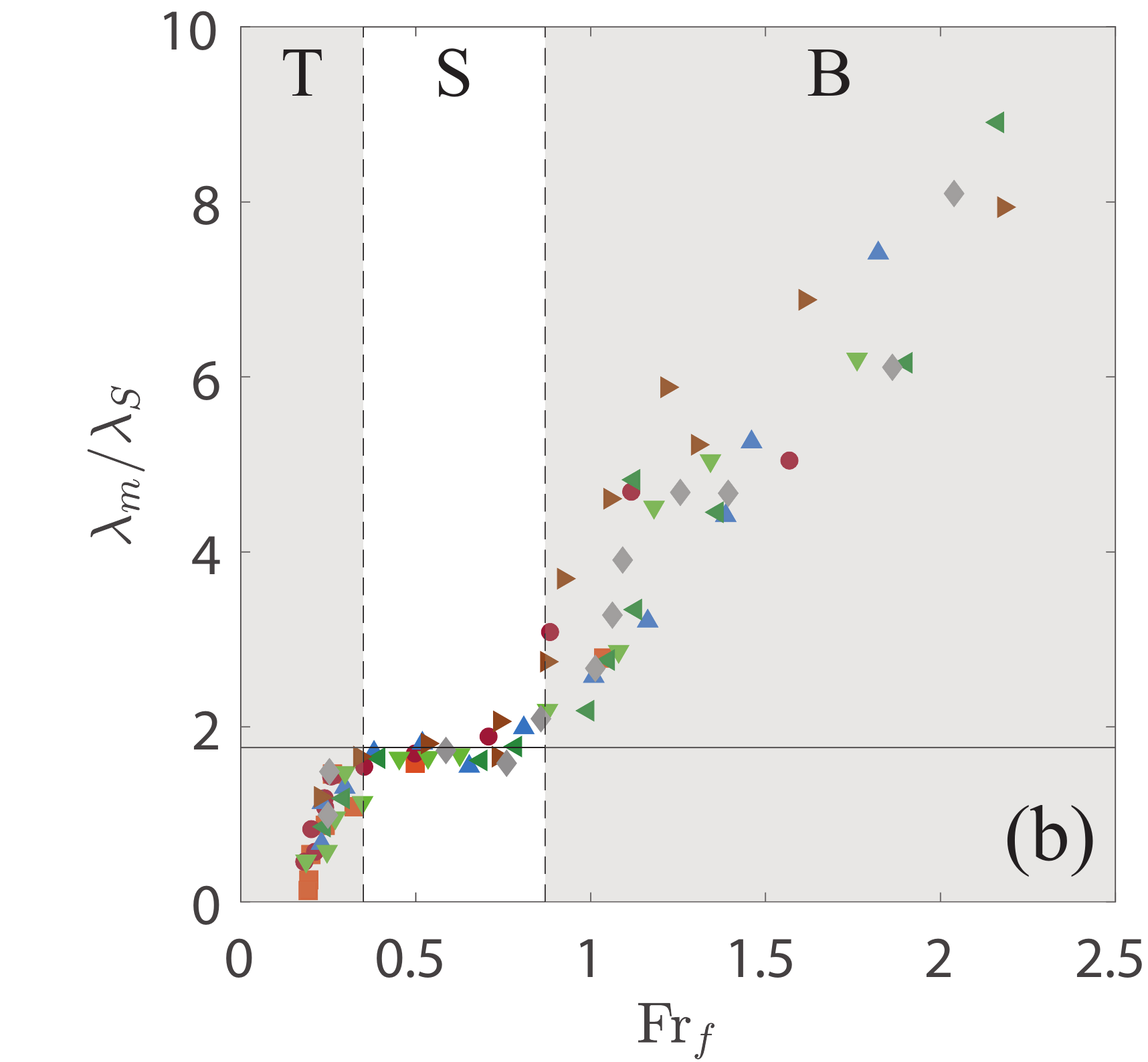}}
  \caption{(a) Plot of $A^*(1+A^*/2)/(1+A^*)$ as a function of $\mathrm{Fr}_{f}$ for all experiments, with $A^*=A_m/h_0$. The best linear fit (full line) of slope 0.95 for the data corresponding to the solitary wave regime ($ 0.35 \lesssim \mathrm{Fr}_{f} \lesssim 0.87$) is also shown. (b) Plot of $\lambda_m/\lambda_S$ as a function of $\mathrm{Fr}_f$, with the expected plateau value $\lambda_m/\lambda_S=1.76$ (horizontal full line).}
\label{fig:fig_V}
\end{figure}

\noindent Figure \ref{fig:fig_V}(a) shows the rescaled experimental data using the left-hand side of equation (\ref{eqArelSWint}), with $A_S=A_m$, as a function of $\mathrm{Fr}_f$. All the experimental data for $ 0.35 \lesssim \mathrm{Fr}_f \lesssim 0.87$ collapse well onto a straight line of slope 0.95. Therefore, in the corresponding experiments, the advancing granular front acts as a moving piston with an effective maximal velocity $v_{f_m} = ~~0.95\ v_p$. This value is a little smaller than one, which may come from the fact that the granular front is not solid but porous and not perfectly vertical.  The explicit expression for $A_S/h_0$ as a function of $\mathrm{Fr}_p$ is obtained as the only positive solution of the second-order equation (\ref{eqArelSWint}):

\begin{equation}
\frac{A_S}{h_0}=\mathrm{Fr}_p + \sqrt{1+{\mathrm{Fr}_p}^2}-1.
\label{eqArelSW}
\end{equation}

\noindent The red dashed curve in figure \ref{fig:fig_III}(a) corresponds to equation (\ref{eqArelSW}) with $\mathrm{Fr}_p~=~0.95~\ \mathrm{Fr}_f$, and fits well the data for $ 0.35 \lesssim \mathrm{Fr}_f \lesssim 0.87$.  The exact expression (\ref{eqArelSW}) can be approximated, using Taylor series, by $A_S/h_0 = \mathrm{Fr}_p + {\mathrm{Fr}_p}^2/2$, which differs by less than 4\% up to $\mathrm{Fr}_p \simeq 0.87$. Note that the expected transition from solitary waves to bores can simply be inferred from the intercept of the approximate law $A_S/h_0 = \mathrm{Fr}_p + {\mathrm{Fr}_p}^2/2$ for solitons with the approximate linear law $A_B/h_0 \simeq  \sqrt{2} \, \mathrm{Fr}_p$ for bores. Considering $A_S = A_B$, by continuity, leads to the critical Froude number ${\mathrm{Fr}_p}_c = 2(\sqrt{2}-1) \simeq 0.8$ and thus to the critical local Froude number ${\mathrm{Fr}_f}_c \simeq 2(\sqrt{2}-1)/0.9 \simeq 0.9$. This value corresponds well to the observed transition value of ${\mathrm{Fr}_f} \simeq  0.87$.
Figure \ref{fig:fig_V}(b) reports the ratio of the experimental mid-height width $\lambda_m$ relative to the expected length $\lambda_S$ obtained from equation (\ref{eqKdVtheory}). We observe a clear plateau value $\lambda_m/\lambda_S \simeq 1.8$ for $ 0.35 \lesssim \mathrm{Fr}_f \lesssim 0.87$, in agreement with the expected value  $2 \cosh^{-1} ( \sqrt{2} ) \simeq 1.76$. This further confirms that the waves observed in regime S correspond to solitary waves. The red dashed curve in figure \ref{fig:fig_III}(b), corresponding to the theoretical prediction obtained by combining equations (\ref{eqKdVtheory}) and (\ref{eqArelSW}), fits well the data with $\mathrm{Fr}_p \simeq 0.95 \, \mathrm{Fr}_f$, and can be approximated by $ 2 \cosh^{-1} ( \sqrt{2} ) \lambda_S/h_0 \simeq 2.03 \, (1-\mathrm{Fr}_p/2) \, {\mathrm{Fr}_p}^{-1/2}$.
In this regime of solitary waves, the relative width $\lambda_m/h_0$ of the wave decreases for increasing $\mathrm{Fr}_f$ in contrast with the other wave regimes, which makes possible a clear differentiation between them. Hence, while equations (\ref{eqArelHJ}) and (\ref{eqArelSW}) give almost the same predictions for the relative amplitude in regime S, as observed in figure \ref{fig:fig_III}(a), only the solitary wave model estimates correctly the width of the waves in this regime. From these observations, we conclude that solitary waves are generated for a narrow range of $\mathrm{Fr}_f$, where the granular collapse acts as a piston wavemaker with variable velocity. Note that these solitary waves are not stable but break for $\mathrm{Fr}_f \gtrsim 0.60$, corresponding to  $A_m/h_0 \gtrsim 0.73$, not far from the critical value 0.78 predicted by \citet{1986_tanaka}.

\begin{figure}
  \centering
  {\includegraphics[width=0.45\linewidth]{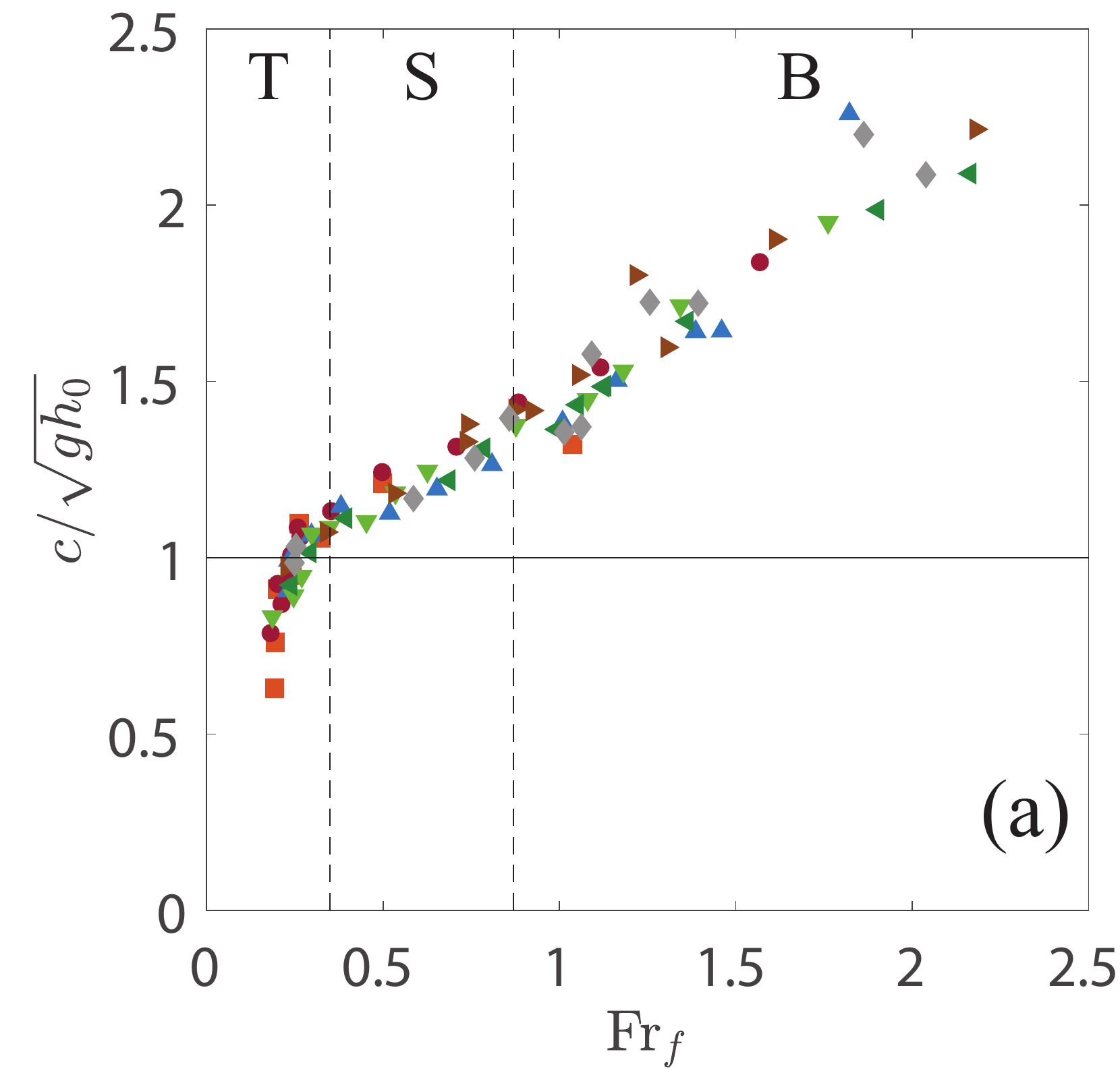}}
  {\includegraphics[width=0.45\linewidth]{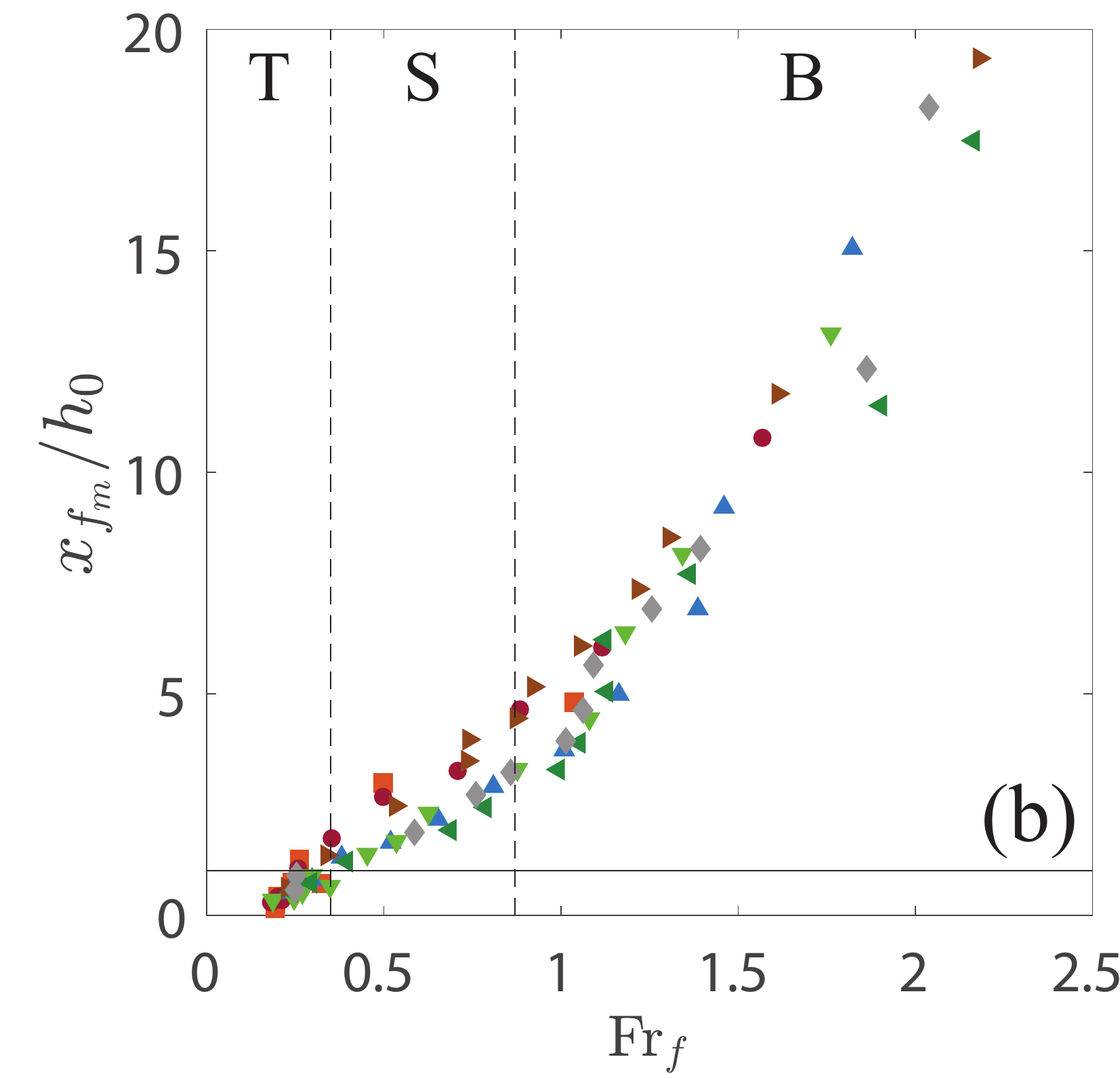}}
  \caption{(a) Evolution of $c/\sqrt{gh_0}$ with $\mathrm{Fr}_f$, where the horizontal full line emphasizes a ratio value of one. (b) Plot of $x_{f_m}/h_0$ as a function of $\mathrm{Fr}_f$, where the horizontal full line shows a ratio value of one.}
\label{fig:fig_VI}
\end{figure}

\subsection{Non-linear transition waves}
\label{subsec:3.1}

For small values of the local Froude number (regime T, with $\mathrm{Fr}_f \lesssim 0.35$), the waves are characterized by strong unsteadiness (wave flattening after generation), and the presence of a dispersive minor wave train. These waves recall one of those observed by \citet{2013_viroulet} and correspond to the non-linear transition waves reported by \citet{2004_fritz}. Figure \ref{fig:fig_III} shows that both the amplitude and wavelength of the generated waves vary abruptly at $\mathrm{Fr}_f \simeq 0.2$. Hence the local Froude number $\mathrm{Fr}_f$, based on the horizontal velocity of the granular front relative to the velocity of shallow-water waves, is not the relevant parameter in regime T any more, for two reasons. First, the shallow-water condition begins to break down in this regime, as observed in figure \ref{fig:fig_VI}(a), where the evolution of the dimensionless wave velocity $c/\sqrt{gh_0}$ is reported as a function of $\mathrm{Fr}_f$. While $c/\sqrt{gh_0} > 1$ when $\mathrm{Fr}_f \gtrsim 0.35$ for regimes S and B, $c/\sqrt{gh_0}$ decreases abruptly below one when $\mathrm{Fr}_f \lesssim 0.35$ in regime T. Second, in figure \ref{fig:fig_VI}(b), where the maximum horizontal extension of the granular front $x_{f_m}$, made dimensionless with the water depth $h_0$, is reported as a function of $\mathrm{Fr}_f$, we observe that $x_{f_m}/h_0 > 1$ when $\mathrm{Fr}_f \gtrsim 0.35$ for regimes S and B,  whereas $x_{f_m}/h_0 < 1$ when $\mathrm{Fr}_f \lesssim 0.35$ for regime T. In this last regime, $x_f$ may not be the only relevant parameter encoding the initial perturbation that generates the wave. This could be the vertical extension $y_f$ of the granular collapse just below the water surface or a combination of $y_f$ and $x_f$. Hence, regime T corresponds to a transition from shallow to deep water conditions, for which developing a theoretical framework is challenging. Note that for deep water conditions, all the parameters characterizing both the granular collapse ($x_f$ and $v_f$) and the generated waves ($A_m$ and $\lambda_m$) are not expected to depend on $h_0$ any more.  In that case, wave generation would be related to the Cauchy-Poisson problem, where a surface impulse generates a modulated wave train \citep{1957_stoker}, which may correspond to the fourth regime of weakly non-linear oscillatory waves mentioned by \citet{2004_fritz}.

\section{Conclusion}
\label{sec:4}
In this study, we reported three different regimes of nonlinear surface waves generated by the gravity-driven collapse of a granular column into water, depending on the local Froude number $\mathrm{Fr}_f$ based on the velocity of the advancing granular front at the water surface. Transient bore waves are observed at high $\mathrm{Fr}_f$ where the granular front acts as a vertical piston pushing the water over the entire water depth and along a distance much larger than $h_0$, at a constant velocity. Solitary waves are observed when $\mathrm{Fr}_f$ is moderate, where the granular front also acts as a vertical piston, but with a varying velocity pushing water along a smaller distance. The amplitude and the mid-height width of the wave generated in these two strongly nonlinear regimes are captured by models derived from shallow-water equations. A third regime corresponding to nonlinear transition waves is observed at low $\mathrm{Fr}_f$, which corresponds to a transition from shallow- to deep-water conditions. In this regime, the local Froude number is no longer the relevant parameter to describe the amplitude and the mid-height width of the wave, and a different model should be developed in future studies. It would also be interesting to provide a better understanding of the dynamics of the grains entering into water in this complex situation \citep{saingier2021falling}, to develop a fully predictive model from the initial granular column to the generated waves.\\

\noindent \textbf{Acknowledgments.} The authors are grateful to J.~Amarni, A.~Aubertin, L.~Auffray and R.~Pidoux for the elaboration of the set-up, and P.-Y. Lagrée, E. Lajeunesse, F. Marin, N. Pavloff and D. Ullmo for fruitful discussions.\\\\
\textbf{Declaration of interests.} The authors report no conflict of interest.

\bibliographystyle{jfm}
\bibliography{jfm}

\end{document}